\documentclass[12pt]{article}

\usepackage{mathrsfs}
\usepackage{amsfonts}
\usepackage{chet}

\preprint{UCSD-PTH-11-14}

\title{Scale without Conformal Invariance:\\\vspace{6pt}
Theoretical Foundations}

\author{Jean-Fran\c{c}ois Fortin, Benjam\'\i{}n Grinstein and Andreas
Stergiou\emails{jffortin@physics.ucsd.edu, bgrinstein@ucsd.edu,
stergiou@physics.ucsd.edu}}

\affiliation{Department of Physics, University of California, San Diego, La
Jolla, CA 92093 USA}

\abstract{We present the theoretical underpinnings of scale without
conformal invariance in quantum field theory.  In light of our results the
gradient-flow interpretation of renormalization-group (RG) flow is
challenged, due to deep connections between scale-invariant theories and
recurrent behaviors in the RG.  We show that, on scale-invariant
trajectories, there is a redefinition of the dilatation current that leads
to generators of dilatations that generate dilatations.  Finally, we
develop a systematic algorithm for the search of scale-invariant
trajectories in perturbation theory.}

\date{July 2011}

\begin{document}

\maketitle

\begin{center}
  \begin{minipage}{0.8\textwidth}
    \begin{center}
      \textbf{Note Added}
    \end{center}
    \vspace{-10pt}
    This paper examines aspects of theories with scale but without
    conformal invariance, quoting the result of~\cite{Fortin:2011ks} as an
    example of such a theory. While the original interpretation of the
    result of~\cite{Fortin:2011ks} was incorrect, as was later realized by
    the authors~\cite{Fortin:2012hn}, the theoretical treatment of scale
    without conformal invariance and the features of such theories
    presented in this work are correct.
  \end{minipage}
\end{center}

\newsec{Introduction}

It has long been presumed that, under mild assumptions, scale invariance
implies conformal invariance in relativistic quantum field theory.
Although no proof is known in $d>2$ spacetime dimensions, until very
recently \rcite{Fortin:2011ks} a credible counterexample was lacking.

In $d=2$ spacetime dimensions, Polchinski \rcite{Polchinski:1987dy},
following an argument of Zamolodchikov \rcite{Zamolodchikov:1986gt}, proved
that scale invariance implies conformal invariance for unitary quantum
field theories with finite correlation functions.  The technical
assumptions of unitarity and finiteness of the correlation functions play
an important role in the proof, and counterexamples which do not satisfy
these assumptions have been found.  Indeed, a scale-invariant model without
conformal invariance, in which correlation functions of the energy-momentum
tensor are not finite, was discovered by Hull and Townsend
\rcite{Hull:1985rc}.  Moreover, the theory of elasticity in $d=2$ Euclidean
dimensions, a non-reflection-positive theory, was shown by Cardy and Riva
to display scale but not conformal invariance \rcite{Riva:2005gd}.  Other
counterexamples are also known
\rcite{Iorio:1996ad,Ho:2008nr}---nevertheless unitarity or finiteness of
the correlation functions is violated in each case.\foot{For related work
see also Ref.~\rcite{Nakayama:2009fe, Nakayama:2010wx, Nakayama:2010zz,
Jackiw:2011vz, ElShowk:2011gz, Antoniadis:2011gn}.}

Polchinski also showed that a unitary scale-invariant theory of scalar
fields in $d=4-\epsilon$ is automatically conformally invariant at one-loop
order \rcite{Polchinski:1987dy}.  The argument he used is simple: for
couplings $g_i$ with beta functions $\beta^i$, scale invariance implies
$\beta^i=\mathcal{Q}^i$ with $\mathcal{Q}^i\neq0$, while conformal
invariance requires $\mathcal{Q}^i=0$ (see Eqs.~\eqref{SolnScale}).
Therefore, if by direct computation one shows that
$(\mathcal{Q}^i)^*\beta^i=0$, then scale implies conformal invariance.
Later on, his result was extended (also at one-loop order) to a theory of
scalar fields and Weyl fermions, in $d=4-\epsilon$, by Dorigoni and Rychkov
\rcite{Dorigoni:2009ra}.  Recently, we showed that for a unitary quantum
field theory of scalar fields exclusively scale implies conformal
invariance to two loops, while in a unitary quantum field theory of Weyl
fermions and no more than one real scalar field, scale invariance implies
the vanishing of the beta functions (and hence conformal invariance) to all
orders in the loop expansion \rcite{Fortin:2011ks}.

More surprisingly, though, we showed that, in more general (unitary)
theories, scale invariance does not necessarily imply conformal invariance
beyond the one-loop order; the Polchinski--Dorigoni--Rychkov argument
breaks down at two loops. For example, the ``$y^3\lambda$'' two-loop term
in the Yukawa beta functions leads to an obstruction to the
Polchinski--Dorigoni--Rychkov argument.  This term also generates an
impediment to expressing the renormalization group~(RG) flow as a gradient
flow as shown by Wallace and Zia \rcite{Wallace:1974dy}, and a deep
connection between scale-invariant trajectories and recurrent behaviors is
revealed.

In this paper we will investigate the theoretical consequences of scale
without conformal invariance for four-dimensional quantum field theories of
vector fields, scalar fields and Weyl fermions, and we will discover
remarkable properties of scale-invariant theories.  As we will argue,
scale-invariant trajectories correspond to rare RG flows, namely recurrent
behaviors (i.e., limit cycles and ergodic behavior).  Specific well-defined
examples of scale-invariant trajectories in $d=4-\epsilon$ (which are
unitary, with finite correlation functions and energy bounded from below)
and in $d=4$ (unfortunately with energy unbounded from below) will be
discussed elsewhere. Here we content ourselves with reviewing the
systematic expansion we used in \rcite{Fortin:2011ks} to search for
recurrent flows.

The paper is organized as follows: section~\ref{SvC} reviews the conditions
under which a theory is scale and/or conformally invariant.  The conditions
for scale without conformal invariance are then translated to conditions on
the beta functions with the help of the new improved energy-momentum
tensor.  An analysis of the RG flow of scale-invariant trajectories leads
to a connection with RG recurrent behaviors.  Such behaviors imply that RG
flows are not gradient flows, and interesting consequences, e.g., for the
$a$-theorem, are discussed.  Finally, an investigation of dilatation
generators for scale-invariant theories allows one to conclude that
dilatation generators do generate dilatations for scale-invariant theories,
even with non-vanishing beta functions.  The implications of scale without
conformal invariance on correlation functions are also briefly discussed.
Section~\ref{ScaleInvTrajectories} describes a general technique to
discover scale-invariant trajectories that are not conformal in generic
quantum field theories.  Finally, we conclude in section \ref{Conclusion}.

\newsec{Scale versus Conformal Invariance}[SvC]

\subsec{Preliminaries}

Let us first review under which circumstances a quantum field theory is
scale or conformally invariant
\rcite{Callan:1970ze,Coleman:1970je,Polchinski:1987dy}.  The most general
form of the dilatation current is
\eqn{\mathcal{D}^\mu(x)=x^\nu T_\nu^{\phantom{\nu}\!\mu}(x)-V^\mu(x)\,,
}[DilatationCurrent]
where $T^{\mu\nu}(x)$ is any symmetric energy-momentum tensor and
$V^\mu(x)$, the virial current,\foot{Strictly speaking the ``field virial''
is a very specific current defined, e.g., in Eq.~(A.14) of
Ref.~\rcite{Coleman:1970je}.  We are relaxing the strict interpretation of
the term.} is any local operator that does not explicitly depend on
$x^\mu$.  The former is determined by the spacetime nature of scale
transformations, while the latter, an internal transformation, contributes
to the scaling dimensions of the fields of the theory.  Notice that the
allowed freedom in the choice of the symmetric energy-momentum tensor is
balanced by the liberty to arbitrarily select the current $V^\mu(x)$.
Since it is finite and not renormalized, the new improved energy-momentum
tensor $\Theta^{\mu\nu}(x)$ \rcite{Callan:1970ze} will be a particularly
helpful choice of $T^{\mu\nu}(x)$ in the following.

For any given choice of energy-momentum tensor, the dilatation current will
be conserved and the theory will exhibit scale invariance if there exists a
virial current such that
\eqn{T_\mu^{\phantom{\mu}\!\mu}(x)=\partial_\mu V^\mu(x)\,.  }[DilInv]
For $d>2$ the theory will also feature conformal symmetry if the virial
current is the sum of a conserved current, $J^\mu(x)$, and the divergence
of a two-index symmetric local operator, $L^{\mu\nu}(x)$, such that
\eqn{T_\mu^{\phantom{\mu}\!\mu}(x)=\partial_\mu
V^\mu(x)=\partial_\mu\partial_\nu L^{\mu\nu}(x)\,.  }[ConfInv]
This last statement is equivalent to the existence of a traceless symmetric
energy-momentum tensor \rcite{Polchinski:1987dy}.

Therefore, for a quantum field theory to be scale but not conformally
invariant, it is necessary that Eq.~\DilInv is satisfied, with the
additional requirement that Eq.~\ConfInv is not, i.e., the virial current
is not the sum of a conserved current and the divergence of a two-index
symmetric local operator:
\eqn{ T_\mu^{\phantom{\mu}\!\mu}(x)=\partial_\mu V^\mu(x)\,,\text{ where
}V^\mu(x)\neq J^\mu(x)+\partial_\nu L^{\nu\mu}(x)\text{ with }\partial_\mu
J^\mu(x)=0\,.  }[DilNoConfCond]
The possible choices for the virial current are easily determined, since
its spatial integral must be gauge-invariant and, in $d$ spacetime
dimensions, its scaling dimension must be $d-1$.  In a general $d=4$
renormalizable quantum field theory the use of the new improved
energy-momentum tensor in Eq.~\DilNoConfCond will be particularly useful in
constraining the virial current.

\subsec{Scale Invariance and the New Improved Energy-Momentum Tensor
\texorpdfstring{\rcite{Callan:1970ze}}{[CCJ]}}

Using dimensional regularization, the most general classically
scale-invariant $d=4-\epsilon$ renormalizable theory involving gauge
fields, $A_\mu^A(x)$, interacting with real scalar fields, $\phi_a(x)$, and
Weyl fermions, $\psi_i(x)$, belonging to arbitrary representations of the
gauge group,\foot{Upper case indices from the beginning of the roman
alphabet are gauge indices for vector fields.  Lower case indices from the
beginning of the roman alphabet are indices in flavor and gauge space for
scalar fields, while lower case indices from the middle are indices in
flavor and gauge space for Weyl spinors.} is described by the Lagrangian
\begin{multline}\label{Lag}
  \mathscr{L}=-\mu^{-\epsilon}Z_A\tfrac{1}{4g_A^2}F_{\mu\nu}^AF^{A\mu\nu}+\tfrac{1}{2}Z_{ab}^{\frac{1}{2}}Z_{ac}^{\frac{1}{2}}D_\mu\phi_bD^\mu\phi_c+\tfrac{1}{2}Z_{ij}^{\frac{1}{2}*}Z_{ik}^{\frac{1}{2}}\bar{\psi}_ji\bar{\sigma}^\mu
  D_\mu\psi_k-\tfrac{1}{2}Z_{ij}^{\frac{1}{2}*}Z_{ik}^{\frac{1}{2}}D_\mu\bar{\psi}_ji\bar{\sigma}^\mu\psi_k\\
  -\tfrac{1}{4!}\mu^\epsilon(\lambda
  Z^\lambda)_{abcd}\phi_a\phi_b\phi_c\phi_d-\tfrac{1}{2}\mu^{\frac{\epsilon}{2}}(yZ^y)_{a|ij}\phi_a\psi_i\psi_j-\tfrac{1}{2}\mu^{\frac{\epsilon}{2}}(yZ^y)_{a|ij}^*\phi_a\bar{\psi}_i\bar{\psi}_j\,,
\end{multline}
where $\lambda_{abcd}$ is totally symmetric in its indices and
$y_{a|ji}=y_{a|ij}$.  For simplicity, gauge-field kinetic mixings are not
considered.  The kinetic terms are defined through
\eqna{ F_{\mu\nu}^A&=\partial_\mu A_\nu^A-\partial_\nu
A_\mu^A+f^{ABC}A_\mu^BA_\nu^C\,,\\
D_\mu\phi_a&=\partial_\mu\phi_a+i\theta_{ab}^AA_\mu^A\phi_b\,,\\
D_\mu\psi_i&=\partial_\mu\psi_a+it_{ij}^AA_\mu^A\psi_j\,, }
where the gauge-group generators $\theta_{ab}^A$ and $t_{ij}^A$ are
Hermitian (the real-scalar-field generators $\theta_{ab}^A$ are also purely
imaginary and antisymmetric).  By gauge invariance the couplings satisfy
\eqna{
\theta_{a'a}^A\lambda_{a'bcd}+\theta_{b'b}^A\lambda_{ab'cd}+\theta_{c'c}^A\lambda_{abc'd}+\theta_{d'd}^A\lambda_{abcd'}&=0\,,\\
\theta_{a'a}^Ay_{a'|ij}+t_{i'i}^Ay_{a|i'j}+t_{j'j}^Ay_{a|ij'}&=0\,.  }
The beta functions are given by vertex corrections plus wavefunction
renormalizations:
\eqna{ \beta_A&=-\frac{dg_A}{dt}=\gamma_Ag_A\quad\text{(no sum)}\,,\\
\beta_{abcd}&=-\frac{d\lambda_{abcd}}{dt}=-(\lambda\gamma^\lambda)_{abcd}+\gamma_{a'a}\lambda_{a'bcd}+\gamma_{b'b}\lambda_{ab'cd}+\gamma_{c'c}\lambda_{abc'd}+\gamma_{d'd}\lambda_{abcd'}\,,\\
\beta_{a|ij}&=-\frac{dy_{a|ij}}{dt}=-(y\gamma^y)_{a|ij}+\gamma_{a'a}y_{a'|ij}+\gamma_{i'i}y_{a|i'j}+\gamma_{j'j}y_{a|ij'}\,.
}[betafunctions]
Here $\gamma^\lambda$ and $\gamma^y$ are the $\lambda_{abcd}$ and
$y_{a|ij}$ vertex anomalous dimensions computed from the vertex corrections
$Z^\lambda$ and $Z^y$ respectively, while $\gamma_A$, $\gamma_{ab}$ and
$\gamma_{ij}$ are the gauge-field, real-scalar-field and Weyl-fermion
anomalous-dimension matrices computed from the wavefunction
renormalizations $Z_A$, $Z_{ab}^{\frac{1}{2}}$ and $Z_{ij}^{\frac{1}{2}}$,
respectively.  The RG time is defined as $t=\ln(\mu_0/\mu)$.

In this perturbative setting the most general non-trivial candidate for the
virial current is
\eqn{
V^\mu(x)=Q_{ab}\phi_aD^\mu\phi_b-P_{ij}\bar{\psi}_ii\bar{\sigma}^\mu\psi_j\,,
}[CandidateV]
where $Q_{ab}$ is antisymmetric and $P_{ij}$ is anti-Hermitian, i.e.,
$Q_{ba}=-Q_{ab}$ and $P^*_{ji}=-P_{ij}$.  By gauge invariance the unknown
coefficients $Q_{ab}$ and $P_{ij}$ must satisfy
\eqna{ \theta_{a'a}^AQ_{a'b}+\theta_{b'b}^AQ_{ab'}&=0\,,\\
-t_{ii'}^AP_{i'j}+t_{j'j}^AP_{ij'}&=0\,.  }
The unknown coefficients in Eq.~\CandidateV are to be determined by
satisfying Eq.~\DilNoConfCond, which is greatly simplified once the new
improved energy-momentum tensor is used.  One reason for that is that the
new improved energy-momentum tensor $\Theta_{\mu\nu}(x)$ is finite and not
renormalized \rcite{Callan:1970ze}.

From the Lagrangian \eqref{Lag} the trace of the new improved
energy-momentum tensor is given by \rcite{Robertson:1991jk}
\begin{multline*}
  \Theta_\mu^{\phantom{\mu}\!\mu}(x)=\tfrac{\beta_A}{2g_A^3}F_{\mu\nu}^AF^{A\mu\nu}+\gamma_{aa'}D^2\phi_a\phi_{a'}-\gamma_{i'i}^*\bar\psi_ii\bar\sigma^\mu
  D_\mu\psi_{i'}+\gamma_{ii'}D_\mu\bar\psi_ii\bar\sigma^\mu\psi_{i'}\\
  -\tfrac{1}{4!}(\beta_{abcd}-\gamma_{a'a}\lambda_{a'bcd}-\gamma_{b'b}\lambda_{ab'cd}-\gamma_{c'c}\lambda_{abc'd}-\gamma_{d'd}\lambda_{abcd'})\phi_a\phi_b\phi_c\phi_d\\
  -\tfrac{1}{2}(\beta_{a|ij}-\gamma_{a'a}y_{a'|ij}-\gamma_{i'i}y_{a|i'j}-\gamma_{j'j}y_{a|ij'})\phi_a\psi_i\psi_j+\text{h.c.}\,.
\end{multline*}
Recall that a transformation is a symmetry of the theory if the
infinitesimal transformation changes the Lagrangian by a total derivative
\textit{without} the use of the equations of motion (EOMs).  However, the
fact that a current is conserved is determined \textit{with} the help of
the EOMs \rcite{Coleman:1970je}.  Therefore, the dilatation current,
$\mathcal{D}^\mu(x)$, the divergence of which is
\begin{multline*}
  \partial_\mu\mathcal{D}^\mu(x)=\tfrac{\beta_A}{2g_A^3}F_{\mu\nu}^AF^{A\mu\nu}+(\gamma_{aa'}+Q_{aa'})D^2\phi_a\phi_{a'}-(\gamma_{i'i}^*+P_{i'i}^*)\bar\psi_ii\bar\sigma^\mu
  D_\mu\psi_{i'}+(\gamma_{ii'}+P_{ii'})D_\mu\bar\psi_ii\bar\sigma^\mu\psi_{i'}\\
  -\tfrac{1}{4!}(\beta_{abcd}-\gamma_{a'a}\lambda_{a'bcd}-\gamma_{b'b}\lambda_{ab'cd}-\gamma_{c'c}\lambda_{abc'd}-\gamma_{d'd}\lambda_{abcd'})\phi_a\phi_b\phi_c\phi_d\\
  -\tfrac{1}{2}(\beta_{a|ij}-\gamma_{a'a}y_{a'|ij}-\gamma_{i'i}y_{a|i'j}-\gamma_{j'j}y_{a|ij'})\phi_a\psi_i\psi_j+\text{h.c.}\,,
\end{multline*}
is conserved if
\eqna{ \beta_A&=0\,,\\
\beta_{abcd}&=-Q_{a'a}\lambda_{a'bcd}-Q_{b'b}\lambda_{ab'cd}-Q_{c'c}\lambda_{abc'd}-Q_{d'd}\lambda_{abcd'}\,,\\
\beta_{a|ij}&=-Q_{a'a}y_{a'|ij}-P_{i'i}y_{a|i'j}-P_{j'j}y_{a|ij'}\,,
}[SolnScale]
for the remaining terms vanish with the help of the EOMs.\foot{Note that,
at the quantum level, the divergence of a current has three types of
contributions: the usual contribution proportional to the EOMs, the
contribution which corresponds to the classical violation of the symmetry,
and a possible anomaly.  Here the classical violation of the symmetry
vanishes, while the combination of the anomaly and the virial current is
proportional to the EOMs.}  The form of Eqs.~\SolnScale is unmodified to
all orders.  Moreover, since operators related to the EOMs are finite and
not renormalized \rcite{Politzer:1980me,Robertson:1991jk}, the dilatation
current, and consequently the virial current, is finite and not
renormalized for scale-invariant theories.  Furthermore, in $d=4$ spacetime
dimensions the virial current has dimension exactly $3$, although it is not
conserved---the theory is scale but not conformally invariant.\foot{This
argument is consistent with the different unitarity bounds for conformal
versus scale-invariant field theories \rcite{Grinstein:2008qk}.}  Notice
that, in terms of the vertex corrections, the conditions for scale
invariance imply
\eqna{
(\lambda\gamma^\lambda)_{abcd}&=(\gamma_{a'a}+Q_{a'a})\lambda_{a'bcd}+(\gamma_{b'b}+Q_{b'b})\lambda_{ab'cd}+(\gamma_{c'c}+Q_{c'c})\lambda_{abc'd}+(\gamma_{d'd}+Q_{d'd})\lambda_{abcd'}\,,\\
(y\gamma^y)_{a|ij}&=(\gamma_{a'a}+Q_{a'a})y_{a'|ij}+(\gamma_{i'i}+P_{i'i})y_{a|i'j}+(\gamma_{j'j}+P_{j'j})y_{a|ij'}\,.
}[VertexAnomDim]
Thus, for scale invariance to hold, the vertex anomalous dimensions must
have very specific forms.  This fact will play an important role later: it
will quantum-mechanically generate the scaling dimensions required in the
dilatation generator.

Finally, it is easy to see from Eq.~\DilNoConfCond that a theory is
scale-invariant but not conformally invariant, if and only if there is a
non-trivial virial current such that $\partial_\mu V^\mu(x)\neq0$ and
Eqs.~\SolnScale or \VertexAnomDim are satisfied.

\subsec{Renormalization Group Flow along Scale-invariant Trajectories}

A quantum field theory with a non-trivial scale-invariant trajectory is a
theory with a solution to Eq.~\DilNoConfCond, or equivalently
Eqs.~\SolnScale or \VertexAnomDim.\foot{Of course, conformal fixed points
still satisfy Eqs.~\SolnScale and \VertexAnomDim with trivial virial
current.}  As mentioned in the introduction, the
Polchinski--Dorigoni--Rychkov argument breaks down at two-loop order in a
theory with enough scalars and fermions.  Hence, scale invariance does not
necessarily imply conformal invariance.  Once a non-trivial scale-invariant
solution $(g_A,\lambda_{abcd},y_{a|ij})$ has been found, it is easy to
solve exactly the RG equations on the scale-invariant trajectory.  Indeed,
with the RG time defined as $t=\ln(\mu_0/\mu)$, the RG evolution on a
scale-invariant trajectory is remarkably simple; it is
\eqna{ \bar{g}_A(t)&=g_A\,,\\
\bar{\lambda}_{abcd}(t)&=\widehat{Z}_{a'a}(t)\widehat{Z}_{b'b}(t)\widehat{Z}_{c'c}(t)\widehat{Z}_{d'd}(t)\lambda_{a'b'c'd'}\,,\\
\bar{y}_{a|ij}(t)&=\widehat{Z}_{a'a}(t)\widehat{Z}_{i'i}(t)\widehat{Z}_{j'j}(t)y_{a'|i'j'}\,,
}[RGSoln]
where the $\widehat{Z}(t)$ matrices are given by
\eqna{ \widehat{Z}_{aa'}(t)&=\big(e^{Qt}\big)_{aa'}\,,\\
\widehat{Z}_{ii'}(t)&=\big(e^{Pt}\big)_{ii'}\,.  }[RGevolCoupling]
Notice that any other point
$(\bar{g}_A(t,g,\lambda,y),\bar{\lambda}_{abcd}(t,g,\lambda,y),\bar{y}_{a|ij}(t,g,\lambda,y))$
on the scale-invari\-ant trajectory will satisfy Eqs.~\SolnScale, since the
couplings and, by the gauge transformations of $Q_{ab}$ and $P_{ij}$, the
beta functions and, also, the anomalous dimensions, transform homogeneously
along the scale-invariant trajectory:
\eqna{
\bar{\beta}_{abcd}(t)&=\widehat{Z}_{a'a}(t)\widehat{Z}_{b'b}(t)\widehat{Z}_{c'c}(t)\widehat{Z}_{d'd}(t)\beta_{a'b'c'd'}\,,\\
\bar{\beta}_{a|ij}(t)&=\widehat{Z}_{a'a}(t)\widehat{Z}_{i'i}(t)\widehat{Z}_{j'j}(t)\beta_{a'|i'j'}\,,\\
\bar{\gamma}_{ab}(t)&=\widehat{Z}_{a'a}(t)\widehat{Z}_{b'b}(t)\gamma_{a'b'}\,,\\
\bar{\gamma}_{ij}(t)&=\widehat{Z}_{i'i}^*(t)\widehat{Z}_{j'j}(t)\gamma_{i'j'}\,.
}[RGSolnbeta]
Here, unbarred parameters are evaluated at $(g_A,\lambda_{abcd},y_{a|ij})$,
i.e., at the scale-invariant solution.  The behavior \RGSolnbeta ensures
that $Q_{ab}$ and $P_{ij}$ are constant along the scale-invariant
trajectory.  Indeed, from Eqs.~\SolnScale and \RGSoln one can see that
trajectories with $Q_{ab}$ and/or $P_{ij}$ that are functions of RG time
are not possible, for these trajectories would then intersect trajectories
with constant $Q_{ab}$ and $P_{ij}$.  Such intersecting trajectories cannot
occur in well-posed initial value problems.

It is perhaps surprising that scale-invariant trajectories are full-fledged
RG trajectories and not simply points since they are always referred to as
such in the literature.  However, because scale without conformal
invariance implies the non-vanishing of at least one of the beta functions,
if a scale-invariant solution $(g_A,\lambda_{abcd},y_{a|ij})$ exists,
complete scale-invariant trajectories must exist and are described by
Eqs.~\RGSoln and \RGevolCoupling.  Note that because $\widehat{Z}_{ab}(t)$
is orthogonal and $\widehat{Z}_{ij}(t)$ unitary, if one of the beta
functions is non-zero at some point on a scale-invariant trajectory, then
at least one of the beta functions will be non-zero at any other point on
the trajectory, i.e., the theory always flows.

Defining matrices $\widetilde{Z}(t)$ as
\eqna{ \widetilde{Z}_{aa'}(t)&=\big(e^{-(\gamma^\phi+Q)t}\big)_{aa'}\,,\\
\widetilde{Z}_{ii'}(t)&=\big(e^{-(\gamma^\psi+P)t}\big)_{ii'}\,,
}[RGevolAnomDim]
the vertex corrections and wavefunction renormalizations on scale-invariant
trajectories are simply
\eqn{\begin{aligned}
Z_{abcd,a'b'c'd'}^\lambda&=\big(\widehat{Z}^{-1}\widetilde{Z}\widehat{Z}\big)_{aa'}\big(\widehat{Z}^{-1}\widetilde{Z}\widehat{Z}\big)_{bb'}\big(\widehat{Z}^{-1}\widetilde{Z}\widehat{Z}\big)_{cc'}\big(\widehat{Z}^{-1}\widetilde{Z}\widehat{Z}\big)_{dd'}\,,
& Z_{aa'}^{\frac{1}{2}}&=\big(\widetilde{Z}\widehat{Z}\big)_{aa'}\,,\\
Z_{a|ij,a'|i'j'}^y&=\big(\widehat{Z}^{-1}\widetilde{Z}\widehat{Z}\big)_{aa'}\big(\widehat{Z}^{-1}\widetilde{Z}\widehat{Z}\big)_{ii'}\big(\widehat{Z}^{-1}\widetilde{Z}\widehat{Z}\big)_{jj'}\,,
 & Z_{ii'}^{\frac{1}{2}}&=\big(\widetilde{Z}\widehat{Z}\big)_{ii'}\,.
\end{aligned}}[VertexCorrWavefunction]
With the help of Eqs.~\RGevolCoupling and \RGevolAnomDim, it is a
straightforward computation to check that Eqs.~\VertexCorrWavefunction lead
to scale-invariant trajectories as described by Eqs.~\SolnScale.

\subsec{Scale Invariance and Recurrent Behaviors}

There is a close connection between theories with scale but without
conformal invariance and recurrent RG behaviors.  Indeed, since $Q_{ab}$ is
real antisymmetric and $P_{ij}$ anti-Hermitian, they both have purely
imaginary eigenvalues.  In other words, the matrices \eqref{RGevolCoupling}
are elements of $SO(N_S)\times U(N_F)$ which correspond to the symmetry
group of the kinetic terms of the $N_S$ real scalars and $N_F$ Weyl
fermions.  Consequently, scale-invariant trajectories described by
Eqs.~\RGSoln must be periodic or quasi-periodic.  In other words,
scale-invariant trajectories exhibit limit cycles or ergodicity, with
oscillation frequencies determined by the eigenvalues of the virial current
\rcite{Wilson:1970ag,Wilson:1973jj,Lorenz:1963yb,TUS:TUS136}.

The connection between scale invariance and recurrent behaviors can be
understood intuitively.  Indeed, RG evolution is related to the dilatation
current \DilatationCurrent, which is a combination of a spacetime scale
transformation and an internal transformation of the fields.  Since the
internal transformation is a transformation in the compact group
$SO(N_S)\times U(N_F)$, the field transformations must eventually rotate
back to the identity or arbitrary close to the identity.  Thus, since RG
translations are generated by scale transformations, and scale
transformations are related to internal transformations by the conserved
dilatation current, RG evolution along scale-invariant trajectories that
are not conformal must return to, or arbitrarily close to, the starting
point.  Hence, scale-invariant trajectories are periodic or quasi-periodic.
Notice that the converse is also true: limit cycles and RG trajectories
exhibiting ergodic behavior are scale-invariant trajectories with no
conformal symmetry, since physical properties on an RG trajectory are
independent of RG time.  However, it is possible that scale invariance on
such a trajectory is broken to a discrete subgroup.\foot{The physics lore
that limit cycles and ergodicity imply perpetual oscillations in the
scattering cross sections is thus correct \rcite{Wilson:1970ag}.  Close to
a scale-invariant trajectory, the scattering cross sections $\sigma(s)$ in
terms of the center-of-mass energy $s$ oscillate, i.e., $s\sigma(s)=c(s)$
with $c(s)$ a periodic or quasi-periodic function.  The scattering cross
sections obey the standard scaling law, with $c(s)$ a constant, only for
theories approaching conformal fixed points.}  These interesting RG
behaviors have far-reaching consequences as discussed in the next section.

At this point some might be puzzled by the argument of the previous
paragraph.  Indeed, since RG running on a scale-invariant trajectory can be
seen as a field redefinition, one might be tempted to argue that
scale-invariant trajectories are really fixed points.  However, the
appropriate field redefinition is RG-time-dependent and consequently it
generates an RG flow as advocated, e.g., in Ref.~\rcite{Wegner:1974aa}.  In
other words, although the RG flow can be interpreted as a field
redefinition, it is impossible to make all beta functions vanish on a
scale-invariant trajectory.\foot{Note that all exact RG flows can be
obtained by RG-time-dependent field redefinitions, see e.g.,
\cite{Latorre:2000jp}.  Moreover, since wavefunction renormalization
operators are redundant, it is necessary for scale invariance that the
beta-function operators are redundant on scale-invariant trajectories---the
beta functions can then be absorbed in the anomalous dimensions as
discussed in section~\ref{Dilatations}.}

Finally, note that recurrent behaviors are $n$-dimensional compact
subspaces of coupling space where $n=1$ for limit cycles and $n>1$ for
ergodic behaviors.  Although a complete analysis of the behavior of RG
trajectories near scale-invariant trajectories has not been undertaken, at
first sight it seems like an extreme amount of fine-tuning is necessary for
a theory to exhibit limit cycles (exactly as in theories that sit at a
fixed point).  The prospect does not appear as grim for ergodicity.  Since
the compact subspace is completely spanned (in infinite RG time), any UV
theory defined in the compact subspace will display ergodicity.  A careful
analysis of RG trajectories near scale-invariant trajectories would shed
light on the character (attractive, repulsive, etc.) of scale-invariant
trajectories.

\subsec{Scale-invariant Theories, Gradient Flows and the
\texorpdfstring{$a$}{a}-theorem}

It has long been known that recurrent behaviors such as limit cycles and
ergodicity imply that RG flows are not gradient flows
\rcite{Wallace:1974dy}.  An RG flow is a gradient flow if the beta
functions, $\beta^i=-\tfrac{dg^i}{dt}$, can be written as
\eqn{\beta^i(g)=G^{ij}(g)\frac{\partial c(g)}{\partial g^j},}
with $G_{ij}(g)$ a positive-definite metric,
$G^{ik}G_{kj}=\delta_{\phantom{i}\!j}^i$, and $c(g)$ a function of the
couplings $g^i$.  Along an RG trajectory, the potential $c(g(t))$ is a
monotonically decreasing function,
\eqn{\frac{dc(g(t))}{dt}=-G_{ij}(g)\beta^i\beta^j\leq0,}
Clearly, scale-invariant trajectories cannot be produced by gradient flows.

The obstruction appears first at two loops, which is consistent with the
literature \rcite{Wallace:1974dy,Osborn:1989td,Jack:1990eb}.  For example,
it was pointed out in \rcite{Wallace:1974dy} that a specific
\textit{one}-loop contribution to the quartic-coupling beta functions
(schematically $y^4$) and a specific \textit{two}-loop contribution to the
Yukawa beta functions (schematically $y^3\lambda$) arise from the same
$\lambda y^4$ term in an appropriately constructed potential $c(g)$.  Such
an obstruction has important repercussions in the study of RG flows.
Intuitively, a non-trivial RG flow is seen as an irreversible process,
where the high-momentum degrees of freedom are integrated out at large
distances.  In other words, the number of massless degrees of freedom
should always decrease along non-trivial RG flows.  However, on
scale-invariant trajectories, the number of massless degrees of freedom is
constant.  This implies that the ``strongest''  version of the $a$-theorem
\rcite{Barnes:2004jj}, i.e., that RG flows are gradient flows, is wrong.
Note that the ``stronger'' claim---that the potential $c(g)$ is
monotonically decreasing, $dc/dt\leq0$---still stands.  The inequality is
saturated if the theory is scale-invariant.

A supersymmetric example of scale without conformal invariance is still
missing, and so it is possible that scale implies conformal invariance for
supersymmetric theories.  In that case, the ``strongest'' version of the
$a$-theorem might still be valid for supersymmetric theories.

\subsec{Why Dilatation Generators Generate Dilatations}[Dilatations]

The authors of Ref.~\rcite{Coleman:1970je} showed that dilatation
generators do not generate dilatations (in non-scale-invariant quantum
field theories).  They demonstrated at low orders that quantum anomalies
can be absorbed into a redefinition of the scaling dimensions of the
fields, but that at high orders this is not possible.  In modern language,
the two anomalies\foot{In Ref.~\rcite{Coleman:1970je} a third type of
anomalies, the Schwinger terms, arose in the Callan--Symanzik equations for
conserved currents.  However, Schwinger terms do not arise in dimensional
regularization and thus can be safely ignored here.} correspond to the
anomalous dimensions and the beta functions.  The former can be safely
absorbed into a redefinition of the scaling dimensions of the fields,
preserving scale invariance, but the latter, generically cannot, thus
breaking scale invariance in the quantum theory. This is made manifest by
the trace anomaly, the statement that once the equations of motions are
applied the trace of the energy momentum tensor vanishes if and only if the
beta-functions vanish.

It is interesting to see how scale-invariant quantum field theories
circumvent the results of Ref.~\rcite{Coleman:1970je}.  At conformal fixed
points the vertex corrections are fixed in terms of the anomalous
dimensions, as can be seen by setting the beta functions to zero in
Eqs.~\betafunctions. By contrast, for a theory to be scale-invariant,
Eqs.~\SolnScale must be satisfied, in which case the vertex corrections
satisfy Eqs.~\VertexAnomDim. Therefore, as in CFTs, the vertex corrections
are fixed, only now they are not given in terms of the anomalous
dimensions, but rather in terms of the generalized anomalous dimensions,
$\gamma+Q$ for scalars and $\gamma+P$ for spinors.  Consequently, on
scale-invariant trajectories it is possible to absorb both the anomalous
dimensions \textit{and} the beta functions into a redefinition of the
scaling dimensions of the fields, thus preserving scale invariance and
leading to dilatation generators that generate dilatations.

To see more precisely how this works, recall that the naive Ward identity
of scale invariance is improved to the Callan--Symanzik equation
\rcite{Coleman:1970je} in the quantum theory.  Indeed, for an arbitrary
collection of fields $\varphi_i(x)$ and couplings $g_i$, defined at the
renormalization scale $M$, the effective action $\Gamma[\varphi(x),g,M]$
satisfies
\eqn{ \left[M\frac{\partial}{\partial M}+\beta^i\frac{\partial}{\partial
g^i}+\gamma_j^{\phantom{j}\!i}\int
d^4x\,\varphi_i(x)\frac{\delta}{\delta\varphi_j(x)}\right]\Gamma[\varphi(x),g,M]=0\,.}
The authors of Ref.~\rcite{Coleman:1970je} pointed out that the
beta-function contributions cannot be absorbed into a redefinition of the
dilatation current, and thus generators of dilatations do not generate
dilatations, except, of course, if the theory is conformal.  However, for
scale-invariant theories the beta functions have very specific linear
dependence on the couplings.  If
$\delta\phi_j=Q^{\phantom{j}\!i}_j\phi^{\phantom{\dagger}}_i$ is an
infinitesimal transformation that is a symmetry of the kinetic terms, then
a formal symmetry-relation is obtained by counter-rotating the coupling
constants, leading to
\eqn{ \left[-Q_j^{\phantom{j}\!i}g^j\frac{\partial}{\partial g^i}
+Q_j^{\phantom{j}\!i}\int d^4x\,\varphi_i(x)\frac{\delta}
{\delta\varphi_j(x)}\right]\Gamma[\varphi(x),g,M]=0\,.}
Then, using Eqs.~\SolnScale it is obvious
that we can substitute the beta-function terms for new
anomalous-dimension-like terms, which lead to a new version of the
Callan--Symanzik equation for the effective action:
\eqn{ \left[M\frac{\partial}{\partial
M}+(\gamma_j^{\phantom{j}\!i}+Q_j^{\phantom{j}\!i})\int
d^4x\,\varphi_i(x)\frac{\delta}{\delta\varphi_j(x)}\right]\Gamma[\varphi(x),g,M]=0\,.}
We have managed to recast the Callan--Symanzik equation as if the beta
functions vanished and the anomalous dimensions were the generalized ones.
This is not to say that the couplings do not flow. Rather, the flow is
precisely such that one can equivalently solve the equations as if the beta
functions vanished but the anomalous dimensions were replaced by the
generalized ones.  Clearly, there is no longer an obstruction to absorbing
the beta-function contributions into a redefinition of the dilatation
current.

It is reassuring that we can arrive at the same conclusion by a different,
largely independent argument, namely by analyzing the Poincar\'e algebra
augmented by dilatation transformations. The beta functions on
scale-invariant trajectories generate the appropriate scaling dimensions
required by the inclusion of the virial current in the dilatation current.
Classically, it is easy to see that the dilatation current
\DilatationCurrent leads to new ``classical'' contributions to the scaling
dimensions of the fields by using the Lie algebra of Poincar\'e and
dilatation transformations.  The Poincar\'e algebra, augmented with the
dilatation charge, $D=\int \!d^3x\,\mathcal{D}^0(x)$, is
\eqna{
[M_{\mu\nu},M_{\rho\sigma}]&=-i(\eta_{\mu\rho}M_{\nu\sigma}-
\eta_{\nu\rho}M_{\mu\sigma}+\eta_{\nu\sigma}M_{\mu\rho}-
\eta_{\mu\sigma}M_{\nu\rho})\,,\\
[M_{\mu\nu},P_\rho]&=-i(\eta_{\mu\rho}P_\nu-\eta_{\nu\rho}P_\mu)\,,\\
[D,P_\mu]&=-iP_\mu\,, }
and the algebra acts on fields\foot{Note that the notion of quasi-primary
fields is vacuous in a scale-invariant theory without conformal invariance,
since the generator of special conformal transformations, $K_\mu$, does not
exist.  Descendants, i.e., operators obtained by the action of the
generator of translations $P_\mu$, are
however well-defined.} $\mathcal{O}_I(x)$ as
\eqna{
[M_{\mu\nu},\mathcal{O}_I(x)]&=-i(x_\mu\partial_\nu-x_\nu\partial_\mu
+\Sigma_{\mu\nu})\mathcal{O}_I(x)\,,\\
[P_\mu,\mathcal{O}_I(x)]&=-i\partial_\mu\mathcal{O}_I(x)\,,\\
[D,\mathcal{O}_I(x)]&=-i(x\cdot\partial+\Delta)\mathcal{O}_I(x)\,,
}[AlgebraAction]
where $\Sigma_{\mu\nu}$ are the appropriate spin matrices and
$\Delta=\Delta^{\rm cl}+\gamma$ are the scale-dimension matrices.

Since the dilatation current is given by Eq.~\DilatationCurrent, we find
(with the help of Eqs.~\AlgebraAction)
\eqna{
[D,\phi_a(x)]=-i(x\cdot\partial+1)\phi_a(x)-iQ_{ab}\phi_b(x)\,,\\
[D,\psi_i(x)]=-i(x\cdot\partial+\tfrac{3}{2})\psi_i(x)-iP_{ij}\psi_j(x)\,,
}
and thus the virial current leads to new ``classical'' contributions to the
classical scaling-dimension matrices as anticipated above: \eqna{
\Delta_{ab}^{\rm cl}&=\delta_{ab}+Q_{ab}\,,\\ \Delta_{ij}^{\rm
cl}&=\tfrac{3}{2}\delta_{ij}+P_{ij}\,.  } In the analysis of the RG flow,
however, these new ``classical'' contributions are introduced at the
quantum level by the beta functions as dictated by Eqs.~\SolnScale or
\VertexAnomDim.  In other words, although from Eq.~\DilatationCurrent the
virial current contributions to the scaling dimensions seem to have a
classical origin, in the renormalization group analysis they are really
generated by quantum corrections, i.e., the beta functions.  Thus, the
scaling dimensions of the fundamental fields on scale-invariant
trajectories are
\eqna{\Delta_{ab}&=\delta_{ab}+Q_{ab}+{\gamma}_{ab}\,,\\
\Delta_{ij}&=\tfrac{3}{2}\delta_{ij}+P_{ij}+{\gamma}_{ij}\,.}[ScaleDim]
Note that the ``classical'' scaling dimensions are constant along the
scale-invariant RG trajectory, while the anomalous dimensions, due to
Eqs.~\RGSolnbeta, are not.  However, the eigenvalues of the scaling
dimensions \ScaleDim are RG-invariant.

We would like to stress here that the beta functions are \textit{not}
shifted away into the scale dimensions of the fields.  Some might argue
that the beta functions on scale-invariant trajectories should naturally be
absorbed into the scale dimensions of the fields, leading to
scale-invariant fixed points instead of scale-invariant trajectories.  But
what does the shift physically mean?  The only possible way to modify the
beta functions without changing the theory is by performing a scheme
change.  If such a scheme change existed, it would imply that there is a
scheme where the new improved energy-momentum tensor is traceless, leading
to conformal invariance.  However, since physics is scheme-independent,
that is clearly impossible.  Furthermore, since the shift is global, it
affects all theory space, transforming scale-invariant trajectories into
scale-invariant ``fixed points'' \textit{and} conformal fixed points into
conformal ``trajectories''.  However, following
Ref.~\rcite{Polchinski:1987dy} no traceless symmetric energy-momentum
tensor exists on these conformal trajectories, contradicting the general
result of Ref.~\rcite{Polchinski:1987dy}.  The shift is thus physically
ill-defined and it is unwarranted to demand vanishing beta functions on
scale-invariant trajectories.

We can also investigate the vacuum structure of the scale-invariant theory
from the tree-level potential.  By inspection of the Lagrangian \eqref{Lag}
we immediately conclude that, if the tree-level scalar potential is bounded
from below, then the theory on a scale-invariant trajectory has a stable
vacuum (i.e., a global minimum) at the origin of field space, with
vanishing energy density.  If the potential is not bounded from below, then
the theory does not have a stable vacuum.  The possibility of flat
directions is not considered.  Due to Eqs.~\RGSoln and the fact that the
matrices \RGevolCoupling are orthogonal/unitary, this statement is
RG-invariant as expected.  The same is true for CP conservation and the
location of the vacuum in field space.  Notice that an analysis of the
effective potential at one loop does not lead to a better understanding of
the theory around the origin of field space.  Indeed, due to the relative
size of the different couplings on scale-invariant trajectories as seen in
Eqs.~\eqref{ExpansionCouplings}, the one-loop contribution to the effective
potential cannot balance the tree-level contribution.  Thus, the apparent
new minimum of the one-loop effective potential, located exponentially
close to the origin, involves large logarithms and lies outside the regime
of validity of the one-loop approximation.  Although it is impossible to
determine if the origin of field space is a true minimum from the effective
potential point of view, the RG analysis strongly suggests that it is.

Finally, it is of interest to study the behavior of correlation functions
of the fields.  It is well-known that scale invariance, along with
invariance under the Poincar\'e group, restricts the form of two-point and
three-point correlation functions of fields
\rcite{DiFrancesco:1997nk}.\foot{Although quasi-primary fields are unique
to conformal field theories, it is still possible in non-conformal
scale-invariant theories to limit the study of correlation functions only
to fields that are not descendants.}  Focusing on scalar fields
$\mathcal{O}_I(x)$ with scaling dimensions $\Delta_I$ for simplicity, the
two-point and three-point correlation functions must be
\eqna{
\langle\mathcal{O}_I(x_1)\mathcal{O}_J(x_2)\rangle&=
\frac{g_{IJ}}{(x_1-x_2)^{\Delta_I+\Delta_J}}\,,\\
\langle\mathcal{O}_I(x_1)\mathcal{O}_J(x_2)\mathcal{O}_K(x_3)\rangle&=
\sum_{\substack{\delta_1+\delta_2+\delta_3=\\
\Delta_I+\Delta_J+\Delta_K}}\frac{c_{IJK}^{\delta_1\delta_2\delta_3}}
{(x_1-x_2)^{\delta_1}(x_2-x_3)^{\delta_2}(x_3-x_1)^{\delta_3}}\,,
}
where $g_{IJ}$ and $c_{IJK}^{\delta_1\delta_2\delta_3}$ are constant.  Note
that, contrary to conformal field theories, two-point correlation functions
of fields with different dimensions do not necessarily vanish for
scale-invariant theories.  Three-point correlation functions are even less
constrained.  Concentrating on fundamental fields, the form of the
two-point functions for real scalar fields can be obtained from the
algebra:
\eqn{
\langle\phi_a(x)\phi_b(0)\rangle=\left[(x^2)^{-\frac{\Delta}{2}}\,
G^\phi\,(x^2)^{-\frac{\Delta^T}{2}}\right]_{ab}\,,
}
where $G_{ab}^\phi$ is a constant real symmetric matrix.

Finally, since the operator product expansion (OPE) already incorporates
scale invariance, no new constraints arise for the OPE on scale-invariant
trajectories.  This is in contrast to conformal theories where very
powerful results can be derived with the use of the generator of special
conformal transformations on the OPE.

\newsec{Scale-invariant Trajectories}[ScaleInvTrajectories]

\subsec{Systematic Approach}

As explained in \rcite{Fortin:2011ks} it is possible to systematically
search for conformal fixed points (where the virial current trivially
vanishes), conformal fixed points with enhanced symmetry (where the virial
current is conserved), scale-invariant trajectories (where the virial
current is non-trivial), and scale-invariant trajectories with enhanced
symmetry (where the non-trivial virial current can be decomposed into a
conserved current and a new non-trivial virial current) in the
weak-coupling regime.  One simply needs to expand in the small parameter
$\epsilon$ the couplings,
\eqn{
g_A=\sum_{n\geq1}g_A^{(n)}\epsilon^{n-\frac{1}{2}},\hspace{1cm}
\lambda_{abcd}=\sum_{n\geq1}\lambda_{abcd}^{(n)}\epsilon^n,
\hspace{1cm}y_{a|ij}=\sum_{n\geq1}y_{a|ij}^{(n)}\epsilon^{n-\frac{1}{2}}\,,
}[ExpansionCouplings]
and the unknown parameters in the virial current,
\eqn{
Q_{ab}=\sum_{n\geq2}Q_{ab}^{(n)}\epsilon^n,\hspace{1cm}P_{ij}=
\sum_{n\geq2}P_{ij}^{(n)}\epsilon^n\,.}[ExpansionVirial]
The form of the expansions \ExpansionCouplings and \ExpansionVirial is
dictated by the beta functions for the couplings \rcite{Jack:1984vj} and by
the Polchinski--Dorigoni--Rychkov argument for the virial current
\rcite{Polchinski:1987dy,Dorigoni:2009ra}.  The only requirement on the
small parameter is to allow a (partial) cancellation of the first and
second non-trivial contributions to the beta functions.  For example, the
small parameter can be the $\epsilon$ of $d=4-\epsilon$, or the specific
function of the number of colors and flavors in a theory of the Banks--Zaks
type in $d=4$ \rcite{Banks:1981nn}.\foot{Notice that, in the large number
of colors and flavors limit, it is more natural to use generalized 't Hooft
couplings \rcite{'tHooft:1973jz} for all couplings in
Eqs.~\eqref{ExpansionCouplings} and \ExpansionVirial.}
It is natural to ask what happens to the scale-invariant trajectories in
the strong-coupling regime.  Once the non-perturbative effects are large,
all confidence in the expansion above is lost, and it is impossible to
argue for the existence of scale-invariant trajectories.  However, one can
imagine that there are scale-invariant trajectories that have sections both
in the perturbative and the non-perturbative regime.  It is also likely
that the scale-invariant trajectories survive in an intermediate regime as
one transitions to strong coupling.  For example for a theory in
$d=4-\epsilon$ the RG flows in the $\epsilon\to1$ limit may give
(strongly-coupled) examples of cycles in $d=3$.  Moreover, in a theory of
the Banks--Zaks type one could assume that scale-invariant trajectories
exist in the strong-coupling regime, much as is done for the conjectured
superconformal fixed points in Ref.~\rcite{Seiberg:1994pq}.  However, one
cannot parallel the argument of Ref.~\rcite{Seiberg:1994pq} for the case of
scale-invariant trajectories, since it relies on the full superconformal
symmetry of the fixed points, more specifically on the existence of an
R-symmetry.  The same can be said of supersymmetric scale-invariant
trajectories if such theories exist.  Before concluding, we would like to
comment on the scheme-dependence of beta functions.

\subsec{Scheme-dependence of Beta Functions}[Scheme]

It is well known that only the first two terms in the loop expansion of the
beta function of QCD are scheme-independent, and hence that a scheme exists
for which the beta function consists precisely of those first two
terms\rcite{'tHooft:1977am}.  The situation for models with multiple
couplings is only slightly more complex, but seems to be less well
understood.  This may well have been discussed elsewhere, but we have tried
and failed to find it in the literature.  So we will point out here that,
for multi-coupling beta functions, although the one-loop terms are
scheme-independent, the two- and higher-loop terms are not. However, there
is in general no scheme choice that can shift to zero the two- or
higher-loop contributions to the beta functions.

We will arrange all the coupling constants of the model into one vector
$g_\alpha=(g_1,\ldots,g_N)$, where the entries stand for squares of
Yang--Mills couplings, $g_A^2$ where $A$ runs over the group factors of the
gauge group, single powers of scalar quartic couplings, $\lambda_I$ with
$I$ running over all possible quartic operators, and products of Yukawa
couplings, $y_iy_j$ (taking all Yukawa couplings to be real by separating
real from imaginary parts) with $i$ running over the number of Yukawa
terms.  Then, the loop expansion of the running of all coupling constants
can be written in a unified way:
\eqn{
\mu\frac{dg_\alpha}{d\mu}\equiv\beta_\alpha=b_{\alpha\beta\gamma}^{(1)}g_\beta
g_\gamma+b_{\alpha\beta\gamma\delta}^{(2)}g_\beta g_\gamma
g_\delta+\cdots\,.  }

Consider, next, a re-parametrization of these coupling constants:
\eqn{ \bar{g}_\alpha=g_\alpha+A_{\alpha\beta\gamma}g_\beta
g_\gamma+\cdots\,.  }[reparam]
This can be seen to correspond to a change in renormalization scheme.
Indeed, if the re-parametrization in Eq.~\reparam is introduced into the
renormalized Lagrangian, one can simply absorb it into the coupling
constant renormalization factors $Z_g$.  But this has the effect of
additional finite subtractions, that is, a change in scheme.

In terms of the new coupling constants the beta functions read
\eqn{
\bar{\beta}_\alpha=\bar{b}_{\alpha\beta\gamma}^{(1)}\bar{g}_\beta\bar{g}_\gamma+\bar{b}_{\alpha\beta\gamma\delta}^{(2)}\bar{g}_\beta\bar{g}_\gamma\bar{g}_\delta+\cdots\,.
}
It is now a simple exercise to relate the coefficients of the loop
expansion of one set of couplings and the other:
\twoseqn{ \bar{b}^{(1)}_{\alpha\beta\gamma}&=b_{\alpha\beta\gamma}^{(1)}\,,
}[barubari] {
\bar{b}^{(2)}_{\alpha\beta\gamma\delta}&=b_{\alpha\beta\gamma\delta}^{(2)}+\tfrac{2}{3}[(A_{\alpha\delta\rho}b_{\rho\beta\gamma}^{(1)}-A_{\rho\delta\beta}\bar{b}_{\alpha\rho\gamma}^{(1)})+\text{permutations}]\,,
}[barubarii]
where the permutations are over all indices but $\alpha$.  We immediately
see that the one-loop terms are scheme-independent, while the two-loop
terms are not.  Moreover, we also notice that $A$ does not have enough
parameters to allow us to set to zero the two-loop coefficients \barubarii.
The same holds for all higher-loop coefficients of the beta functions.

In dimensional regularization the two-loop beta functions, when evaluated
on scale-invariant trajectories, are of the order of the three-loop terms,
as will be described in future work. Thus, without the knowledge of the
three-loop beta functions, one cannot argue consistently that a solution
obtained using the two-loop beta functions corresponds to a scale-invariant
recurrent flow rather than a fixed point. It should be noted that generally
there does not exist a scheme in which the beta functions are two-loop
exact.  One thus concludes that a three-loop computation is necessary to
establish the existence of scale-invariant trajectories.  These issues will
be investigated in great detail in another publication.

\newsec{Discussion and Conclusion}[Conclusion]

It has long been thought that scale invariance implies conformal invariance
in unitary quantum field theory.  A proof in $d=2$ spacetime dimensions was
known, but no such proof existed for $d>2$ spacetime dimensions.  In this
work we lay down the theoretical foundations for $d=4-\epsilon$ and $d=4$
unitary quantum field theories with finite correlation functions that are
scale but not conformally invariant.

On scale-invariant trajectories the dilatation current is conserved, the
virial current has dimension exactly $3$, although it is not conserved
(something conformal symmetry does not allow), and the RG evolution is
known precisely.  Moreover, scale-invariant trajectories exhibit recurrent
behaviors (limit cycles or ergodicity) with non-vanishing beta functions.
This fact implies that RG flows are not gradient flows and, therefore, the
``strongest'' version of the $a$-theorem is violated.  Finally, dilatation
generators do generate dilatations on scale-invariant trajectories, since
the beta functions can also be absorbed into a redefinition of the scaling
dimensions of the fields.  Indeed, the beta functions generate the
appropriate scaling dimensions required by the non-trivial contribution of
the virial current to the dilatation current.  As expected, statements such
as boundedness of the scalar potential, CP conservation, and location of
the minimum in field space are RG-invariant for scale-invariant theories.
Note, however, that an analysis of the effective potential around the
origin of field space lies outside the range of validity of the one-loop
approximation.

Several explicit counterexamples which display scale without conformal
invariance will be exhibited elsewhere.  Such examples allow the study of
the implications of scale invariance, without the added constraint of
conformal invariance.  They also shatter all hopes for a generic proof that
scale implies conformal invariance in arbitrary spacetime dimensions.  The
perturbative analysis moreover suggests that scale-invariant theories that
are not conformal are generic.

Due to their presumed non-existence, theories with scale but without
conformal invariance have been scantly studied.  For example, it would be
of interest to study the character (attractive, repulsive, etc) of
scale-invariant trajectories in generic quantum field theories.  Possible
phenomenological applications could emerge and result in new ideas for
model building.  Moreover, an analysis in $d>2$ and $d\neq4$ spacetime
dimensions would also shed light on possible violations of the $a$-theorem
in other spacetime dimensions.  Furthermore, a study of scale without
conformal invariance in supersymmetric quantum field theories could also
lead to interesting consequences for the $a$-theorem.  Finally, it would be
interesting to investigate scale-invariant theories in the context of
gauge/gravity duality
\rcite{Awad:1999xx,Awad:2000ac,Awad:2000aj,Nakayama:2011zw}.  We look
forward to addressing some of these questions in the future.

\ack{We thank Antonio Amariti, Ken Intriligator, Pierre Mathieu and Wouter
Waalewijn for useful discussions.  This work was supported in part by the
US Department of Energy under contract DOE-FG03-97ER40546.}

\bibliography{SvCTheory_ref}

\end{document}